\documentstyle[amssymb,aps,prl,twocolumn,epsfig]{revtex}

\def\be{\begin{equation}}
\def\ee{\end{equation}}
\def\bea{\begin{eqnarray}}
\def\eea{\end{eqnarray}}
\def\bma{\begin{mathletters}}
\def\ema{\end{mathletters}}

\begin{document}
\title{Atomic Quantum State Teleportation and Swapping}
\author{A. Kuzmich$^{(1),\dagger }$ and E. S. Polzik$^{(2)}$}
\address{$^{(1)}$Department of Physics and Astronomy, University of Rochester,\\
Rochester, New York 14627 \\
$^{(2)}$Institute of Physics and Astronomy, Aarhus University, 8000 Denmark}
\date{\today }
\maketitle

\begin{abstract}
A set of protocols for atomic quantum state teleportation and swapping
utilizing Einstein-Podolsky-Rosen light is proposed. The protocols are
suitable for collective spin states of a macroscopic sample of atoms, i.e.
for continuous atomic variables. Feasibility of experimental realization for
teleportation of a gas sample of atoms is analyzed.
\end{abstract}

\draft

Quantum teleportation \cite{bennett}, has recently attracted considerable
attention as the means of disembodied transfer of an unknown quantum state.
Besides being a new fundamental concept in quantum physics, it is also
relevant for such applications as quantum computing and quantum memory.
Experimental realizations of quantum teleportation have so far been limited
to teleportation of light \cite{teleexp}, \cite{furusawa}. Since quantum
information processing involves material particles such as atoms and ions,
teleportation of atomic states will be the next important benchmark on the
way to obtaining a complete set of quantum information processing tools.
Several proposals for teleportation of atoms onto atoms and atoms onto light
have been published recently \cite{attelep}.

We propose here a set of protocols for teleportation of a collective spin
state of a macroscopic atomic sample. Our teleportation procedure does not
require strong coupling of a single atom with light, and does not therefore
involve high-Q cavities. We also consider a variation of the procedure that
provides swapping of the spin states of two atomic systems.

As in the e.-m. field quantum teleportation of Refs.\cite{braunstein,furusawa}%
, the protocol for quantum teleportation of spin proposed here relies on the
continuous entanglement of the Einstein-Podolsky-Rosen (EPR) -type output of
an optical parametric oscillator (OPO) below threshold. For parametric gain $%
r$ ($r>0$) the quadratures of the two modes of the EPR state satisfy the
following relations: 
\begin{eqnarray}
\hat{X}_{1}+\hat{X}_{2} &=&\sqrt{2}\exp (-r)\hat{X}_{v},  \nonumber \\
\hat{Y}_{1}-\hat{Y}_{2} &=&\sqrt{2}\exp (-r)\hat{Y}_{v},  \label{quadr}
\end{eqnarray}
where $\hat{X}_{v},\hat{Y}_{v}$ describe a vacuum mode.

As was pointed out in the original teleportation proposal \cite{bennett}, it
is imperative to be able to perform {\it joint} measurements on the quantum
state to be teleported and one of the EPR states. We propose to use
off-resonant atom-photon interaction to achieve this goal. Such interaction
as a probe of the quantum state for atoms with total ground state electronic
angular momentum equal $\hbar /2$ has been analyzed in Ref. \cite{happer},
and recently has been shown to be suitable for spin quantum non-demolition
(QND) measurements \cite{kuz-EPL,klaus,takahashi}, as well as for generation
of entangled samples of atoms \cite{kuz-EPL,kuz-exp}. As shown in Refs.\cite
{happer,kuz-EPL,klaus,takahashi,kuz-exp} the unitary time-evolution operator
corresponding to the interaction with light propagating along the $z$-axis
has the following form: 
\begin{equation}
\hat{U}=\exp {(-ia\hat{S}_{z}\hat{F}_{z})},  \label{QNDHamil}
\end{equation}
where $a$ is given by $a=\frac{\sigma }{AF}\frac{\gamma }{\Delta }\alpha
_{v} $, $\sigma $ is the resonant absorption cross section for an
unpolarized photon on an unpolarized atom of total spin $F$, $A$ is the area
of the transverse cross section of the light beam, $\gamma $ is the
spontaneous emission rate of the upper atomic level, $\Delta $ is the
detuning, ${\bf \hat{F}}=\sum_{\mu }{\bf \hat{F}}^{(\mu )}$ (summed over all
the atoms in the interaction region) is the operator of the collective
ground-state atomic spin, ${\bf \hat{S}}$ is the operator of the Stokes
vector of the optical field integrated over the duration of the interaction.
The dynamic vector polarizability $\alpha _{v}=\pm 1$ for the $D_{1}$
transition of alkali atoms, while $\alpha _{v}=\mp \frac{1}{2}$ for the $%
D_{2}$ transition. Here the upper/lower sign is for the hyperfine sublevel with $F=I\pm %
\frac{1}{2}$, $I $ is the value of the nuclear spin.

Interaction (\ref{QNDHamil}) leads to rotation of the polarization of the
field that is proportional to ${\hat{F}_{z}}$. Subsequent optical
polarization measurements provide the classical information in our
teleportation protocol. In order to do that, each of the two EPR modes
(polarized along the y-axis) is mixed on a polarizing beamsplitter with a strong x-polarized coherent pulse having $%
n_{1}$ and $n_{2}$ photons respectively. Then, for the Stokes vectors of the two resulting bright EPR
fields ${\bf \hat{S}}^{(1)}$ and ${\bf \hat{S}}^{(2)}$ we obtain: 
$\hat{S}_{z}^{(1)} \approx \sqrt{n_{1}}\hat{X}_{1}, \hat{S}_{y}^{(1)}\approx 
\sqrt{n_{1}}\hat{Y}_{1}, 
\hat{S}_{z}^{(2)} \approx \sqrt{n_{2}}\hat{X}_{2}, \hat{S}_{y}^{(2)}\approx 
\sqrt{n_{2}}\hat{Y}_{2}$. That is, the spin components of the Stokes vectors of the bright EPR fields
are given by the corresponding quadrature operators of the OPO output.

{\bf Teleportation of an atomic spin state onto polarization state of light.}
Let Alice pass her bright EPR beam 1 with the Stokes vector ${\bf \hat{S}}%
^{(1)}$ through the cell containing polarized atomic vapor along the z-axis.
The unitary time evolution operator of Eq.(\ref{QNDHamil}) in the case of
small spin fluctuations in ${\bf \hat{F}}$ and ${\bf \hat{S}}$ ($\sqrt{%
\langle (\Delta \hat{F}_{z,y})^{2}\rangle }\ll F$, $\sqrt{\langle (\Delta 
\hat{S}_{z,y})^{2}\rangle }\ll S)$ leads to the following transformations: 
\begin{eqnarray}
\hat{S}_{y}^{(1)(out)} &\approx &\hat{S}_{y}^{(1)(in)}+\frac{an_{1}}{2}\hat{F%
}_{z}^{(in)},  \nonumber \\
\hat{F}_{y}^{(out)} &\approx &\hat{F}_{y}^{(in)}+(aNF)\hat{S}_{z}^{(1)(in)}.
\label{BS}
\end{eqnarray}
These equations for interaction of the collective atomic spin with light
resemble the beamsplitter relations used for teleportation of a mode of
electromagnetic field \cite{braunstein}, \cite{furusawa}. An insightful
analysis of why a linear device of a beamsplitter allows to perform joint
measurements necessary for quantum teleportation has been given by Vaidman
and Yoran \cite{vaidman}. The connection to our atom-light interaction case
takes root in the fact that small rotations of a spin on the Bloch sphere
around some point are equivalent to displacements in the tangent plane of
this point, which can be considered as motions in the phase plane of the
harmonic oscillator.

The {\it nonlinear} atom-light interaction of Eq.(\ref{QNDHamil}) is
analogous to the QND-type interaction for the optical quadratures $\sim \hat{%
X}_{a}\hat{X}_{b}$. Unlike the beamsplitter, which mixes both the $\hat{X}$
and the $\hat{Y}$ quadratures of the two input beams in the same fashion,
interaction $\sim \hat{X}_{a}\hat{X}_{b}\ $mixes only {\it one} pair of
quadratures ($\hat{Y}_{a}$ and $\hat{Y}_{b}$), leaving the other pair ($\hat{%
X}_{a}$ and $\hat{X}_{b})$ unchanged. Relations (\ref{BS}) correspond to a
50:50 beamsplitter when 
\begin{equation}
|a|\sqrt{FNn_{1}/2}=1.  \label{unity-gain}
\end{equation}

Subsequent (destructive) measurement of $\hat{S}_{y}^{(1)(out)}$ provides
Alice with half of the classical information needed for implementation of
''feed-forward'' on Bob's bright EPR beam to complete the quantum
teleportation. The other half must come from the measurement of $\hat{F}%
_{y}^{(out)}$. Although it is possible in principle to use a completely
destructive measurement (e.g., by means of photoionization), experimentally
it may be more convenient to employ another QND measurement using an
auxiliary coherent pulse. Let us rotate the spin by $\pi
/2 $ around the x-axis, so that a pulse propagating along the z-axis will
measure $\hat{F}_{y}^{(out)}$. The corresponding transformation is 
\begin{eqnarray}
\hat{S}_{z}^{(coh)(out)} &\approx &\hat{S}_{z}^{(coh)(in)}+\frac{an_{coh}}{2}%
\hat{F}_{y}^{(out)}  \nonumber \\
&\approx &\hat{S}_{z}^{(coh)(in)}+\frac{an_{coh}}{2}\hat{F}_{y}^{(in)}+\frac{%
n_{coh}}{n_{1}}\hat{S}_{z}^{(1)(in)}.
\end{eqnarray}
The first term is negligible if $n_{coh}/n_{1}\gg 1$. After Alice sends the
results of measurements of $\hat{S}_{y}^{(1)(out)}$ and $\hat{S}%
_{z}^{(coh)(out)}$ to Bob, reconstruction of $\hat{F}^{(in)}$ in ${\bf \hat{S%
}}^{(2)}$ is achieved by Bob rotating the latter around the z-axis by the
angle $\phi _{1}=-\frac{2}{\sqrt{n_{1}n_{2}}}S_{y}^{(1)(out)}$and around the
y-axis by the angle $\phi _{2}=-\frac{2\sqrt{n_{1}}}{n_{coh}\sqrt{n_{2}}}%
S_{z}^{(coh)(out)}$ . The rotations are equivalent to displacements because
the angles are small. We obtain using Eqs.(\ref{quadr}) 
\begin{eqnarray}
\hat{S}_{y}^{(2)(out)} &\approx &\sqrt{\frac{n_{2}}{2FN}}\hat{F}_{z}^{(in)}+%
\sqrt{2n_{2}}\exp (-r)\hat{Y}_{v},  \nonumber \\
\hat{S}_{z}^{(2)(out)} &\approx &\sqrt{\frac{n_{2}}{2FN}}\hat{F}_{y}^{(in)}+%
\sqrt{2n_{2}}\exp (-r)\hat{X}_{v}.  \label{mainresult}
\end{eqnarray}
The last terms in these expressions are due to the extra noise introduced by
the imperfect EPR state. This noise goes to zero as $r$ goes to infinity.
These terms correspond to the
''quantum duty'' of the quantum teleportation \cite{braunstein}, as seen
explicitly in the case of zero parametric gain, $r=0$. Eqs. (\ref{mainresult}%
) show that the Stokes vector ${\bf \hat{S}}_{2}^{out}$ is identical to the
initial collective spin vector of Alice's atoms when 
\begin{equation}
\frac{n_{2}}{2FN}=1.  \label{spin-equal}
\end{equation}

The teleportation protocol described above may serve as a read-out for
quantum memory with the atomic sample as a memory cell. On the other hand, if the
process of mapping of the Stokes vector ${\bf \hat{S}}_{2}^{out}$ onto
another \ atomic spin \cite{map} follows the described atom-to-light
teleportation, atom-to-atom teleportation is achieved.

{\bf Teleportation of atoms onto atoms.} We now describe a protocol which
performs the direct teleportation of the quantum state of Alice's collective
spin onto Bob's collection of atoms without using light as an intermediate
carrier of the quantum state. Suppose we have two macroscopic spin systems
(Alice's and Bob's) in the initial states given by collective spin operators 
$\hat{F}_{A},\hat{F}_{B}$ with mean polarizations $\langle \hat{F}%
_{Ax}\rangle =\langle \hat{F}_{Bx}\rangle $ and with the other two
projections $\hat{F}_{Ay},\hat{F}_{By},\hat{F}_{Az},\hat{F}_{Bz}.$ We also
have at our disposal the EPR source of light as described above, with each
EPR beam mixed with a strong pulse containing equal photon numbers, $%
n_{1}=n_{2}=n$, polarized along the $x$-axis, similar to the previous
section. Alice's EPR beam is phase shifted by $\pi /2$ so that the
Stokes operators are $\hat{S}_{Az}=-\hat{S}_{By},\hat{S}_{Ay}=-\hat{S}_{Bz}$. The protocol begins with Alice sending her EPR beam along the $%
z$-axis and measuring its $y$ -Stokes parameter with the detector $D_{A1}$,
and Bob sending a coherent $x$-polarized pulse $\hat{S}_{B}^{coh}$
containing $n_{c}$ photons along the $y$-axis and measuring its $z$-Stokes
parameter with the detector $D_{B1}$ (see Fig. 1). The resulting atomic states of Alice
and Bob are described by the following operators: 
\begin{eqnarray}
\hat{F}_{Az}^{^{\prime }} &=&\hat{F}_{Az},\hat{F}_{Ay}^{^{\prime }}=\hat{F}%
_{Ay}+\hat{S}_{Az},  \label{atel1} \\
\hat{F}_{By}^{^{\prime }} &=&\hat{F}_{By},\hat{F}_{Bz}^{^{\prime }}=\hat{F}%
_{Bz}-\hat{S}_{By}^{coh}.  \nonumber
\end{eqnarray}
In these equations we combined conditions $n_{1}=n_{2}=n$ and (\ref
{unity-gain}),(\ref{spin-equal}) to obtain unity coupling coefficients
between $\hat{F}$ and $\hat{S}$ projections. The Stokes parameters measured
by detectors $D_{A1}$ and $D_{B1}$ are

\begin{eqnarray}
\hat{d}_{A1} &=&\hat{S}_{Ay}+\hat{F}_{Az},  \label{atel2} \\
\hat{d}_{B1} &=&\hat{S}_{Bz}^{coh}-\frac{n_{c}}{n}\hat{F}%
_{By}\thickapprox -\frac{n_{c}}{n}\hat{F}_{By}
\end{eqnarray}
We assume $n/n_{c}\ll 1$. While the above measurements are performed, the
second EPR beam sent by Alice to Bob begins its journey along the quantum
channel.

\begin{figure}
\epsfig{file=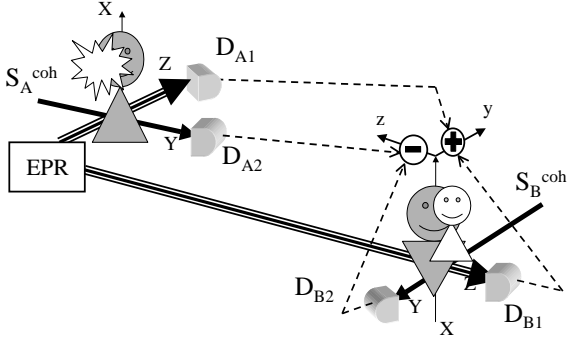,width=3.5in}
\caption{Teleportation of Alice's atomic state onto Bob's state. Alice
communicates with Bob via a quantum channel (EPR beam) and classical
channels (dashed lines). After four measurements with two EPR beams and two
coherent beams ( $S_{A}^{coh},S_{B}^{coh}$) are performed on Alice's atoms
and Bob's atoms with detectors $D_{A1,2},D_{B1,2}$ , Bob's state is shifted
(rotated) along $y$ axis by $d_{A1}+\frac{n}{n_{c}}d_{B1}$ (shown as a plus in a circle)
and along $z$ by $-\frac{n}{n_{c}}d_{A2}-d_{B2}$ (shown as a minus in a circle). As a result
Bob's state becomes Alice's initial state and Alice's state is wiped out.}
\end{figure}

Next Alice sends a coherent $x$-polarized probe containing $n_{c}$ photons
along $y$ axis onto the detector $D_{A2}$. The detector measures

\begin{equation}
\hat{d}_{A2}=\hat{S}_{Az}^{coh}-\frac{n_{c}}{n}\hat{F}%
_{Ay}^{^{\prime }}\thickapprox \frac{n_{c}}{n}\hat{F}_{Ay}^{^{\prime }}  \label{atel3}
\end{equation}

Alice now sends $d_{A1},\frac{n}{n_{c}}d_{A2}$ to Bob along a classical
channel.

When Bob receives the EPR beam from Alice he sends it along $z$ axis onto
the detector $D_{B2}$ which reads

\begin{equation}
\hat{d}_{B2}=\hat{S}_{By}+\hat{F}_{Bz}^{^{\prime }}.  \label{atel4}
\end{equation}

After that the state of Bob's atoms is

\begin{equation}
\hat{F}_{By}^{^{\prime \prime }}=\hat{F}_{By}^{^{\prime }}+\hat{S}_{Bz}, 
\hat{F}_{Bz}^{^{^{\prime \prime }}}=\hat{F}_{Bz}^{^{\prime }}.  \label{atel5}
\end{equation}

To complete the teleportation Bob now rotates his atomic state. We use
displacements instead of rotations to simplify the expressions. Bob's state
is displaced along $z$ by $-\frac{n}{n_{c}}d_{A2}-d_{B2}$ and along $y$ by $d_{A1}+\frac{n}{n_{c}}d_{B1}$
(see Fig.1). The final state of his atoms, according to (\ref{atel2}-\ref{atel5}), is described by the following equations:

\begin{eqnarray}
\hat{F}_{Bz}^{tele}=\hat{F}_{Bz}^{^{\prime \prime }}-\frac{n}{n_{c}}d_{A2}-d_{B2}=-\hat{F}%
_{Ay}, &&  \label{atel7} \\
\hat{F}_{By}^{tele}=\hat{F}_{By}^{^{\prime \prime }}+d_{A1}+\frac{n}{n_{c}}d_{B1}=\hat{F}%
_{Az} &&  \nonumber
\end{eqnarray}
and the teleportation of the unknown state of Alice's collection of
atomic spins onto Bob's atoms is proven (within a rotation of $\pi $ around
the $x$-axis). In the above equations we assumed perfect entanglement
between the EPR beam components ($r\rightarrow \infty $ in Eqs.(\ref{quadr}%
)). 

{\bf Atomic quantum state swapping.} We now describe a
protocol which exchanges initial quantum states of two collections of atomic
spins. Suppose Alice's and Bob's spin samples are in the initial state given by
operators $\hat{F}_{A},\hat{F}_{B}$ with mean polarizations $\langle \hat{F}%
_{Ax}\rangle =\langle \hat{F}_{Bx}\rangle .$ We again have at our disposal
the EPR source of light as described above, with Stokes operators $\hat{S}%
_{z},\hat{S}_{y}$ for one of the beams, and $-\hat{S}_{z},\hat{S}_{y}$ for
the other. Both EPR beams are mixed with strong beams containing equal
photon numbers, $n_{1}=n_{2}=n$, polarized along the $x$ -axis in a way
similar to the previous section. One of the EPR beams is used for a joint
measurement on the $z$ spin components of both samples by sending it through
both of them along the $z$ -axis. The resulting states of atomic samples are:

\begin{eqnarray}
\hat{F}_{Az}^{^{\prime }} &=&\hat{F}_{Az};\hat{F}_{Bz}^{^{\prime }}=\hat{F}%
_{Bz},  \label{2tel1} \\
\hat{F}_{Ay}^{^{\prime }} &=&\hat{F}_{Ay}+\hat{S}_{z};\hat{F}_{By}^{^{\prime
}}=\hat{F}_{By}+\hat{S}_{z};  \nonumber
\end{eqnarray}
and the state of the beam is

\begin{equation}
\widehat{S}_{z}^{^{\prime }}=\hat{S}_{z};\hat{S}_{y}^{^{\prime }}=\hat{S}%
_{y}+\hat{F}_{Az}+\hat{F}_{Bz}.  \label{2tel7}
\end{equation}

In the above equations we assumed conditions (\ref{unity-gain}),(\ref
{spin-equal}) to be fulfilled. Next, the second EPR beam shifted in phase by 
$\pi /2$ is sent along the $y$ -axis of the two atomic samples to perform a
joint measurement on the $y$ spin components. The beam is transformed by the
phase shift and the change of the direction in the following way $-\hat{S}%
_{z},-\hat{S}_{y}\rightarrow \hat{S}_{z}^{r},\hat{S}_{y}^{r}$. The evolution
operator in this new coordinate system is $\hat{U}=\exp {(-ia\hat{S}_{y}^{r}%
\hat{F}_{y}).}$ We obtain

\begin{eqnarray}
\hat{F}_{Az}^{^{\prime \prime }} &=&\hat{F}_{Az}-\hat{S}_{y}^{r}=\hat{F}%
_{Az}+\hat{S}_{y};\hat{F}_{Bz}^{^{\prime \prime }}=\hat{F}_{Bz}+\hat{S}_{y},
\\
\hat{F}_{Ay}^{^{\prime \prime }} &=&\hat{F}_{Ay}^{^{\prime }}=\hat{F}_{Ay}+%
\hat{S}_{z};\hat{F}_{By}^{^{\prime \prime }}=\hat{F}_{By}^{^{\prime }}=\hat{F%
}_{By}+\hat{S}_{z},  \nonumber
\end{eqnarray}
and 
\begin{eqnarray}
\hat{S}_{z}^{r^{\prime }} &=&-\hat{S}_{z}^{^{\prime }}=-\hat{S}_{z}+\hat{F}%
_{Ay}^{^{\prime }}+\hat{F}_{By}^{^{\prime }}=\hat{F}_{Ay}+\hat{F}_{By}+\hat{S%
}_{z},  \label{2tel2a} \\
\hat{S}_{y}^{r^{\prime }} &=&-\hat{S}_{y}^{^{\prime }}=-\hat{S}_{y}. 
\nonumber
\end{eqnarray}
Here we used Eqs. (\ref{2tel1}, \ref{2tel7}). After interacting with the
atoms the two EPR beams are detected by photodetectors $D_{1},D_{2}$. The
Stokes parameter $S_{y}^{^{\prime }}$ (\ref{2tel7}) is measured for the
first beam and the Stokes parameter $S_{z}^{r^{\prime }}$ (\ref{2tel2a}) for
the second. The results of the measurements are used to rotate the atomic
spins in order to achieve the swapping. The projections $\hat{F}%
_{Az}^{^{\prime \prime }}$ and $\hat{F}_{Bz}^{^{\prime \prime }}$ are
displaced by the value $S_{y}^{^{\prime }}$ obtained from $D_{1}$, and the
projections $\hat{F}_{Ay}^{^{\prime \prime }}$ and $\hat{F}_{By}^{^{\prime
\prime }}$ are displaced by the value $S_{z}^{r^{\prime }}$. The results are
$\hat{F}_{Az}^{swap} =\hat{F}_{z}^{^{\prime \prime }}-S_{y}^{^{\prime
}}=-\hat{F}_{Bz},
\hat{F}_{Ay}^{swap} =\hat{F}_{y}^{^{\prime \prime }}-S_{z}^{r^{\prime }}=-
\hat{F}_{By}$. 
Similarly for the other atomic system
$\hat{F}_{Bz}^{swap} =-\hat{F}_{Az}  
\hat{F}_{By}^{swap} =-\hat{F}_{Ay}$ and the initial
quantum states of the two samples have been swapped.

{\bf Feasibility of experimental realization.} The experimental
attractiveness of the above protocols relies on the fact that they can be
applied to the system as simple as a gas of atoms in a cell at room
temperature. The only basic assumption is that the number of atoms in the
teleported ensemble has to be much greater than one. The unity coupling
conditions $\frac{1}{2}an=aNF=1$ yields $n=2FN,|a|=|(\sigma \gamma \alpha
_{v})/(AF\Delta )|=2/n$. The required optical depth of the samples has to
fulfill the condition $\alpha _{\Delta }=\sigma N\gamma /A|\Delta |=\gamma
/|\Delta |\ll 1$ to secure the validity of interaction described by Eq.(\ref
{QNDHamil})$.$ Optimal cross section of the optical beams is then $A=(\sigma
n|\alpha _{v}|\alpha _{\Delta })/(2F)$, that shows that $n\gg 1$ is
necessary to fulfill $A\gg \sigma \approx \lambda ^{2}$ (the condition of
weak focusing, i.e. weak coupling of light to a single atom). If a
collection of, say $N=10^{5},$ $Cs$ atoms in $F=4$ state probed close to the 
$6S_{1/2}\leftrightarrow 6P_{3/2},\gamma =5MHz$ is to be teleported, the
following parameters will be close to optimal: the detuning of probes $%
\Delta =800MHz$, the cell size $10\times 10\times 200\mu m^{3}$, and the
atomic density $n_{A}=5\times 10^{12}cm^{-3}.$ The cell size can vary with
the optimal size scaling roughly as $\sqrt{N}$. The number of photons in EPR
pulses should be around $n=8\times 10^{5}$ with the pulse duration longer
than $100n\sec $ to avoid saturation effects. With a typical finite
bandwidth non-classical light generated by an optical parametric oscillator 
\cite{bandopo} the pulse duration will also have to be longer than the
inverse OPO bandwidth, $\Gamma _{OPO}^{-1}\approx 10n\sec $ to preserve
quantum correlations. A buffer gas can be added into the cell and/or its
walls can be coated to extend the lifetime of the ground state spin state 
\cite{longspin}.

{\bf Summary.} We propose a set of quantum state teleportation and swapping
protocols based on an off-resonant interaction of EPR light with macroscopic
atomic spin ensembles. The three protocols proposed include (a)
teleportation of an atomic state onto a state of light suitable for quantum
memory read out (b) teleportation of an atomic sample onto another atomic
sample and (c) quantum state swapping of two atomic samples. The
macroscopic character of the collective spin teleported unites this proposal
with the previous work on quantum state processing with continuous quantum
variables of light and atoms \cite{furusawa},\cite{braunstein},\cite
{braun-lloyd} and experimental approaches towards creating multiatom
entangled states \cite{kuz-EPL},\cite{kuz-exp},\cite{map}. The absence of
high-Q optical cavities may become an attractive feature simplifying an
experimental realization. One possible realization involving a gas of long
lived spins of alkali atoms is shown to be accessible. Teleportation between
small cells filled with atoms should provide a suitable playground for
studies of quantum memory and continuous quantum information processing.

We gratefully acknowledge useful discussions with B. Julsgaard, I.
Cirac, P. Zoller, K. M\o lmer and I. Walmsley. ESP acknowledges
funding from the Danish Research Council and Thomas B. Thriges Center for
Quantum Information.

${}^{\dagger }$ Present address: NEC Research Institute, Inc., 4
Independence Way, Princeton, NJ 08540-6634.

\end{document}